# OPTIMISATION OF PRESSURE SEWER OPERATION

Jakub Jura, Jan Chyský, Lukáš Novák

Czech Technical University in Prague, Faculty of Mechanical Engineering
Department of Instrumentation and Control Engineering
Technická 4, 166 07 Prague 6
Czech Republic
{jakub.jura; jan.chysky;  lukas.novak} @fs.cvut.cz

Abstract: *The paper deals with the new control method developed for the pressure sewer systems. This method eliminates the disadvantages of currently common used on-off regulation. The major disadvantage is a transition of inconstancies of the effluent production into the sewage system. The propose method is primarily based on the principle of an allocation of the drawing off into the given time slots. This control method is further improved by the extended modules provide higher level of the optimization (learning mode and emergent drawing off).  Proposed method is able to decrease of the standard deviation of pumping even by 80%.*

## 1  Introduction

Pressure sewer is already an established alternative to a conventional gravity sewer (e.g. [1] or [2]). Its main advantage is a considerable independence from the terrain irregularities in which the elements of pressure sewer system are located. The Pressure Sewer System is putted together from the basic house units/stations (schema of one unit is shown at the Figure 2) which are usually bounded up with building. Each mentioned unit has its own septic tank with a grinder pump, which transfers a waste medium (effluent) by the sewer system into a wastewater treatment facility. The local control system is of course a part of this basic unit.

## 2  State of art – on-off regulation and its disadvantages

The pressure sewers are usually controlled by an on-off regulation of the level of effluent in the septic tank (e.g. [3]). The septic tank is equipped by two level sensors – the high sensor that turns on the pump and the low sensor which turns them off. A major disadvantage of this method lies in the constriction of complex problem to the control of an effluent level in the given single tank only – and others important meaningful demands to the control of whole system (containing all units, pipelines, the sewage plant) are not considered. **On-off regulation copies (transmits) inconstancies in the effluent production into the sewage system** (see Figure 1) and is not able to utilize the installed storage capacity of the septic tanks.

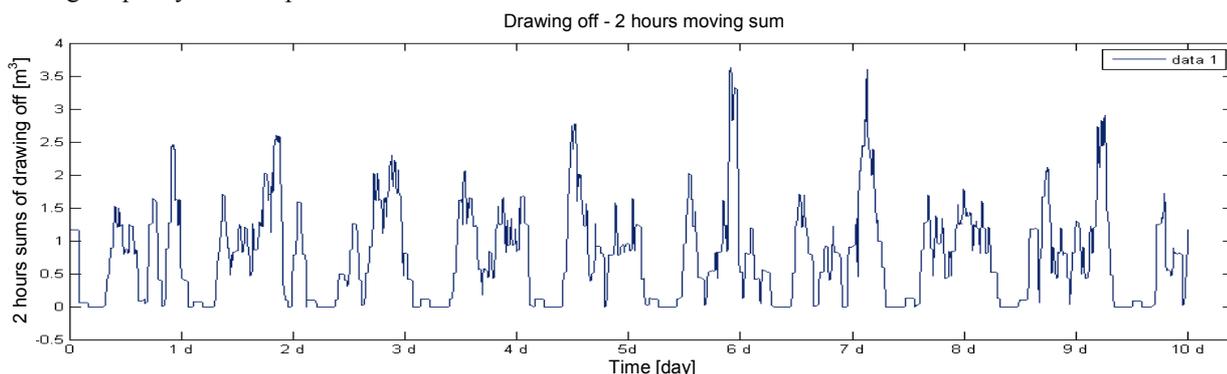

Figure 1: Graph of two hours sums of all drawings off the effluent from a septic tanks into the pressure sewer system. Axis y: two hours sums of all drawing off; Axis x: time (10 days range); Used control mode: **on-off regulation,**

## 3  Aim of optimization

The others requirements to the control of the pressure sewer is **constancy of the load** of sewer pipe network and **steady inflow to the sewage plant**. On-off regulation leads to a very inconstancies and irregular inflow of effluent to the wastewater treatment facility – with significant maximum in the evening peak (e.g. Fig 2 in [4]). The leveling of the wastewater influent to the sewage plant allows its smooth running and especially to reduce the capacity, to which must be the system designed.

First it is possible to design the smaller sewage plant for the same load (output of en effluent). And secondly (and more importantly), it is possible to increase the capacity of the existing pressure sewers system - without its reconstruction (there is not need to dig into the ground!). It is very useful in a situation, when due to further home construction, the new users /producers of effluent/ increased, after the pressure sewer system had been made.

The task – keep up an inflow to the sewage plant relatively constant – is not a trivial task. The effluent production naturally fluctuates and control system can not directly affect them (likewise a flooding). The control system can only use the storage capacity of individual septic tanks intelligently. Moreover, the task to regulate the level of each one septic tank (in its operational limits) has always the highest priority.

## 4 Design of control system

The designed control algorithm [8], which should provide the control functions mentioned above works in a few modes. Each mode is realized by the given control module. Whole control system is a special interconnection of these modules (Figure 5). The system is relatively open and there is possible to add new modules (or disconnect any existing ones).

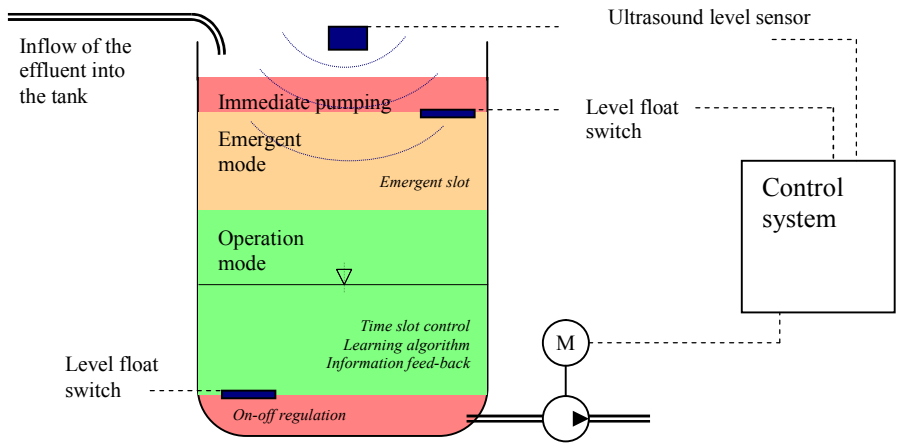

Figure 2: The control of drawing off of the septic tank – the zones related to control modes are illustrated

As the basis of the control algorithm was used old **on-off regulation**, which takes the fail-save role and provides the immediate pumping of the effluent from the septic tank (the red zone - Figure 2). This mode works independently on the others modes.

Operating (or regular) mode is carried out primary by the module of **time slot control** (green zone at Figure 2). This mode provides the time distribution of drawing off up to 24 hours. It levels the daily fluctuations in the wastewater influx to the sewage plant most of all control modes. The principle of the time slot control mode is the allocation of few time slots to each unit/house station. And the drawing off a wastewater can starts in the given allocated time slot only. First, it prevents to uncontrolled amount of stations drawing off simultaneously. And secondly (and it is more important) this mode provides a leveling of daily peaks of effluent production - and consequently more constant inlet wastewater to the sewage system. Similar way of optimization is known from different field of interest – for example from the transportation engineering [5] or from the field of communication network [6].

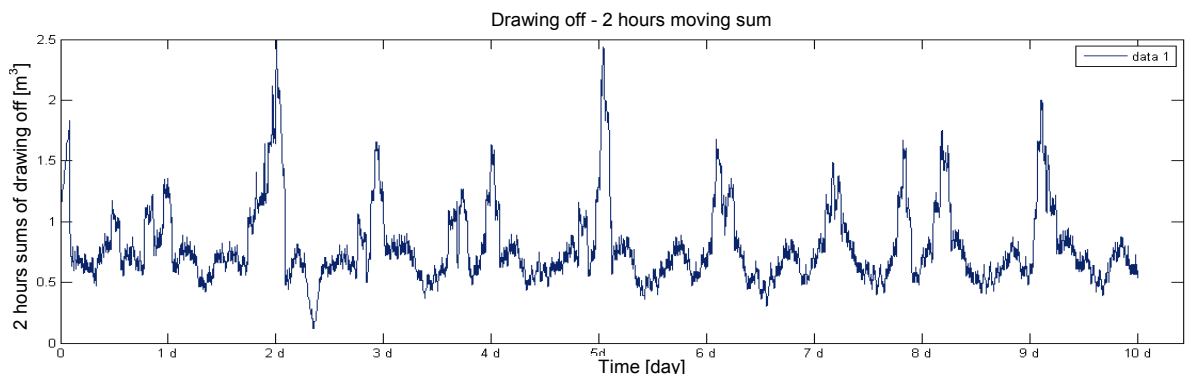

Figure 3: Graph of two hours sums of all drawings off the effluent from a septic tanks into the pressure sewer system. Axis y: two hours sums of all drawing off; Axis x: time (10 days range); Used control mode: **time slot basic**

The next mode which is called **emergent time slot control** (orange area Figure 2) is activated when the level of effluent in the septic tank reaches the warning level. Emergent time slot is the time slot which is similar as a regular one, but is appointed to all stations which exceed the emergent (but not immediate) level. The emergent time slot incoming periodically - for example every tenth time slot – and is designated for emergent drawing off. From the septic tank is not pump all effluent, but only the amount corresponding with length of used time slot. By this attribute the emergent mode differs from the on-off regulation.

The last mode is used for the optimization of both previous time slot control modes (regular and emergent one). This mode is called **learning mode**. The module (D) for this mode modifies (adapts) pumping time in the time slot on the basis of previous production of en effluent. If the level in the septic tank reaches the given value (e.g. orange zone at Figure 2) than the pump time is elongated about given amount of time (e.g. 10 second). And at the other side, if the level of effluent in the septic tank reaches the low level limt (e.g. lower red zone at Figure 2), than the pump time (time of one drawing off) is shorteneed about same amount of time.

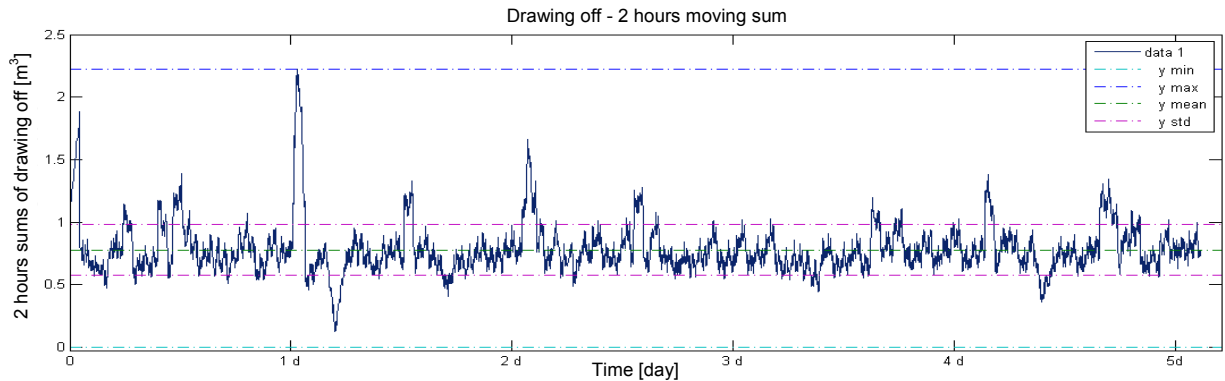

Figure 4: Graph of two hours sums of all drawings off the effluent from a septic tanks into the pressure sewer system. Axis y: two hours sums of all drawing off; Axis x: time (10 days range); Used control mode: **time slot basic + learn**

By the mutual interconnection of all mentioned modules arises the control algorithm providing a) **safety** functions, b) **leveling** the inflow to the sewage plant and c) the high **robustness** of whole system. Individual modules are interconnected together according to the schema at Figure 5.

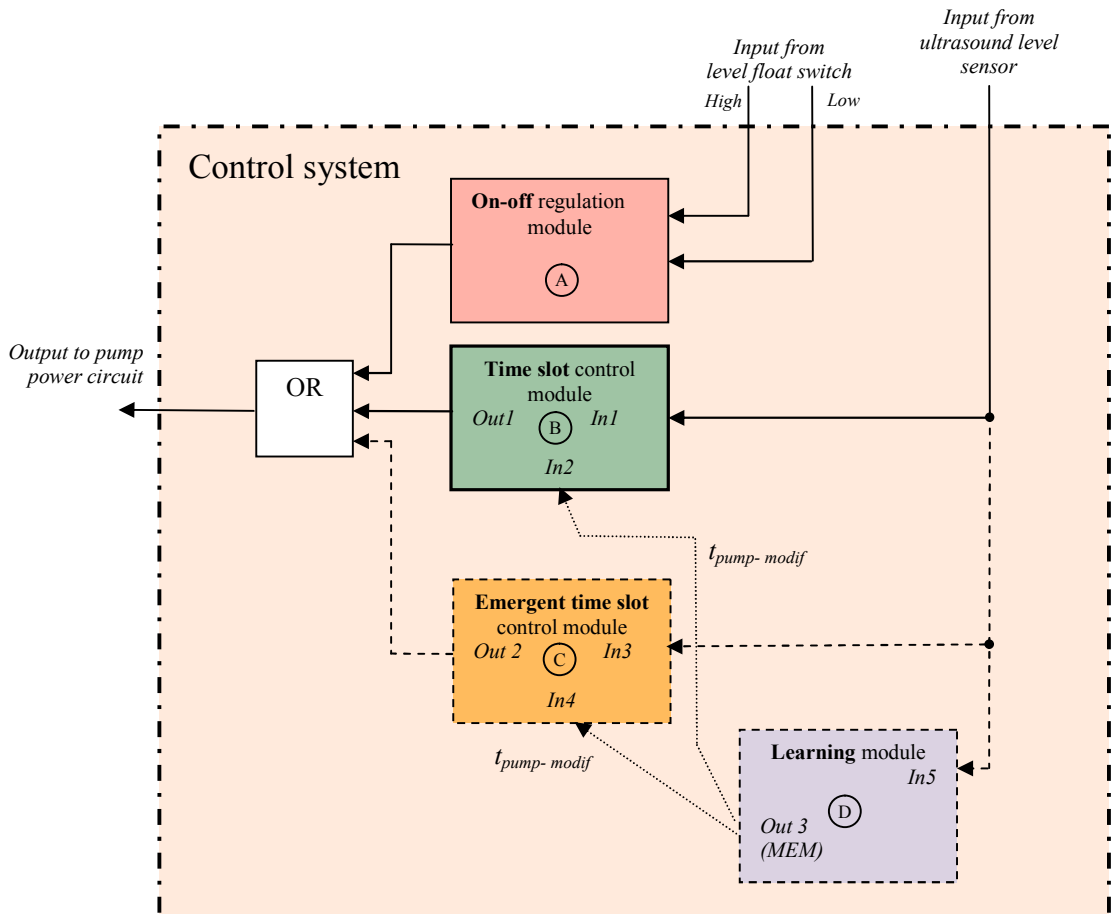

Figure 5: Interconnection of control modules [8]

# 5 Learning module

As it was mentioned above, the main function of the learning module is to fit the basic variable parameter of the proponed control - the length of pumping time in the time slot - to the individual quantity of the waste water production (e.g. by a home unit). In the time slot control mode (B) the average value of production for the given municipality is used. Unhappily this average value has a real variance and for the pressure sewer system it has two nature impacts: 1) time allocated for the pumping of waste water is insufficient and therefore the emergent time slot mode (C) is often used. Lower constancy of the pressure sewer system load is consequent. XOR 2) time allocated for the drawing off is overlarge and the consequence is low level of utilizes of the septic tank capacity for the transfer of the pumping of the waste water into the time periods with lower production (all volume of waste water is drained away from a home septic tank in the firs time slot).

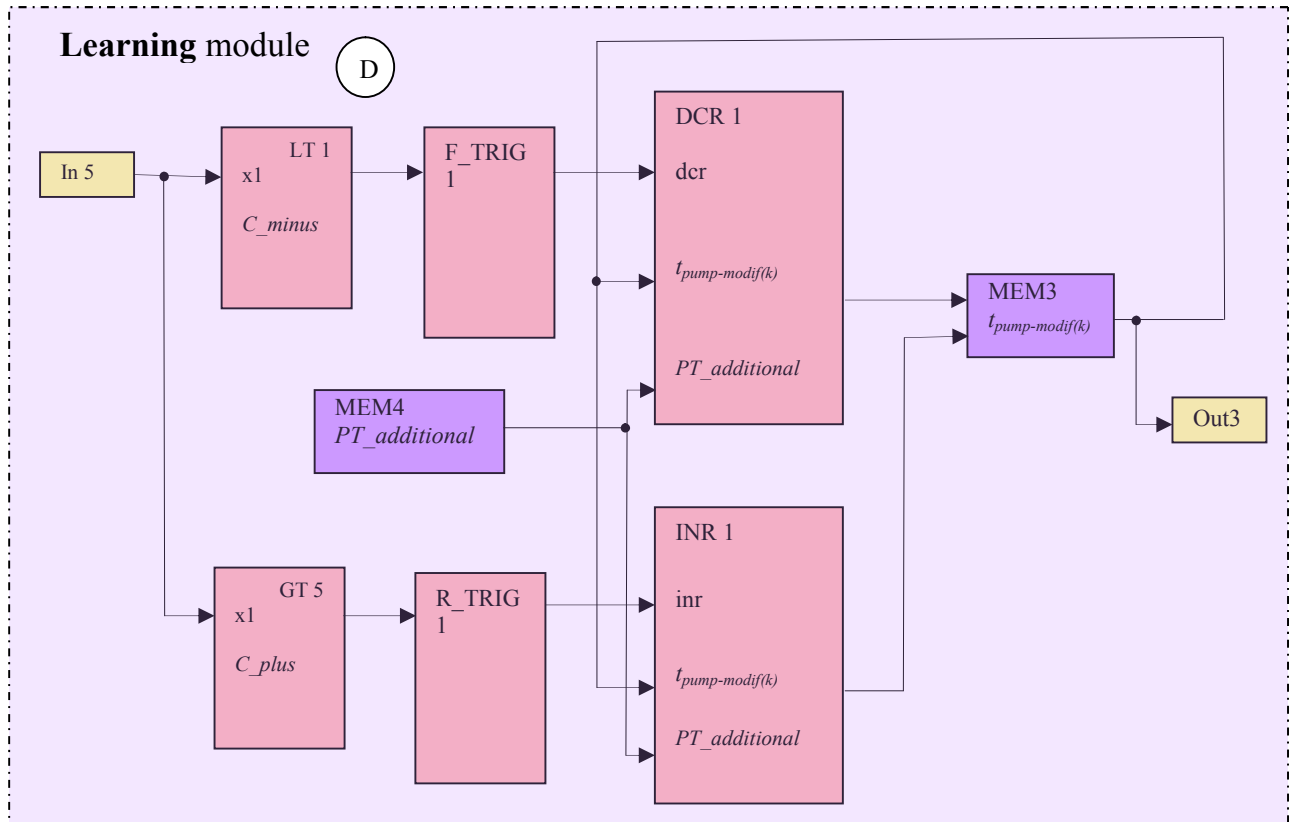

Figure 6: Schema of the Learning module interconnection [8]

The Learning module (D) uses internal memory (MEM 3 on Figure 6) which keep the individual value of the time ($t_{pump-modif(k)}$) about which the basic pumping time is modified for compensation of the variance of the effluent production (basic pumping time is define in module B and C and its length is determine by the mass balance). When the level of the effluent in the septic tank (In 5 is input from the level sensor) falls below the limit C_minus, than the modification pumping time ($t_{pump-modif(k)}$) is shortened by the given time (PT_additional). And conversely - if the level of the effluent in the septic tank reaches up to the limit C_plus, than the modification pumping time ($t_{pump-modif(k)}$) is prolongated by the same time (PT_additional). On the figure
Figure 7 is possible to see the adaptation of the variable ($t_{pump-modif(k)}$) for 12 basic unit (k) to their common production. It is possible to tell, that this module (D) works on the basis of "experience" with the previous length of drawing off from the local sump. The additional time (PT_additional) about which the drawing off is shorter or longer, determines the speed of the learning. The length of this time 10 and 20 seconds was used for numerical simulations and the results are very promising (compare Figure 4 with Figure 1).

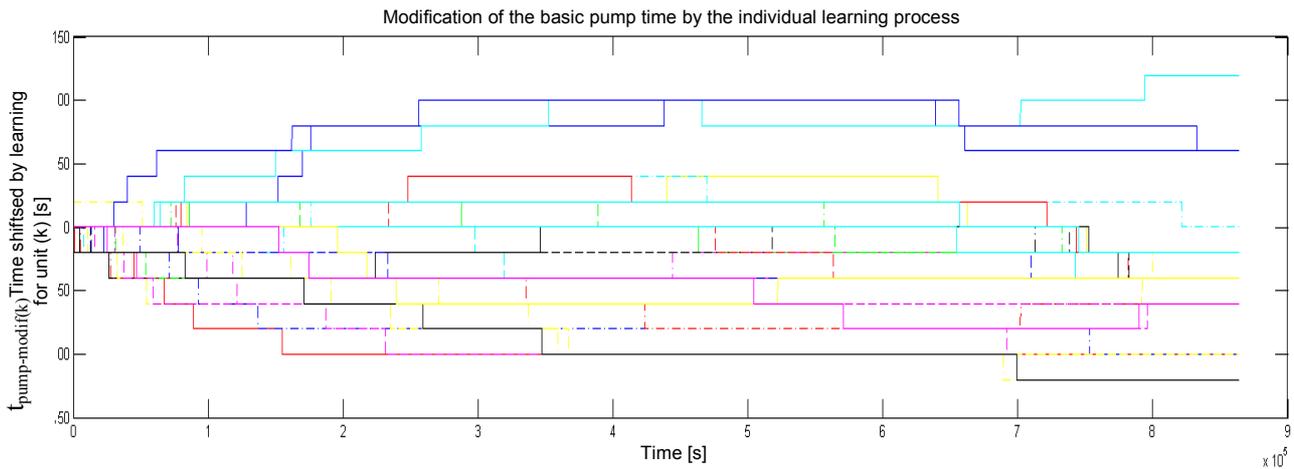

Figure 7: Graph of values of individual modifications of the pump time in the learning mode (D). The adaptation of the pump time depends on the previous experience is possible to see here. Axis y: individual modifications of the pump time; Axis x: time (10 days range); Used control mode: time slot basic + learn

Considering that the high level of robustness of the control method should to be reached, the relatively simple solution was used (see Figure 6). At the other side, whole system consists of big amount of the basic (home) units and each unit works with same algorithm but with different input data. From the methodological point of view this method belongs to the bottom-up methods where solution emerges from a big amount of relatively simple basic unit.

## 6 Conclusion

The impact of single modules to the behavior of whole system was shown separately at figures above (Figure 1, Figure 3 Figure 4). The indicator of load of pressure sewer system was used two hours moving sums of all drawings from septic tanks to the pressure sewer pipelines (on mentioned figures it is possible to see changes in statistical parameters like an average value, standard deviation or maximum and minimum value of observed variables). Although the shapes of the curves which represent courses of drawing off for different configurations of the control system are relatively clear at first glance, theirs numerical characteristics express this difference exactly. At Figure 8 is possible to compare standard deviation of two hours moving sums of all drawings for mentioned configurations of control system.

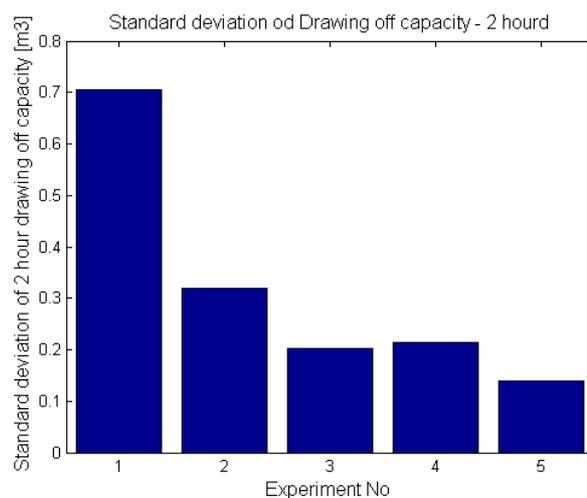

Figure 8: Standard deviation of the 2 hours drawing off capacity. Experiment No:
1 – On-Off regulation (A)
2 –Time slot with FIX pump time (AB)
3- Time slot + learning (ABD)
4 – Time slot with + emergent slot (ABC)
5 – Time slot + learning + emergent slot (ABCD)

The numerical simulation [7] of pressure sewer system operation shows that the proposed control method is able to level the load of pressure sewer system very weightily. The most probably the use of proponed control system [8] will tend to a decrease of a parameters which is need to a design of a wastewater treatment facility like a daily and hourly inequality coefficients or minimal and maximally inflow. Moreover the economic savings appears to be influential in the case of expansion of an insufficient capacity of an existing pressure sewer system (without their reconstruction). At the other side – the use of pressure sewer has an environmental impact. A nowadays massive expansion of pressure sewer systems leads to fast water drainage from the landscape and thus the landscape dehumidifying and dehydrating [10] which continuously leads to small water cycle (SWC) violation and more significant losing of water from the landscape [9]. Local raising the temperature is of course nature impact of dehumidifying. Nevertheless this problem doesn't lie in the pressure sewer but in the business model of their use primarily.

**Acknowledgement:** This work was supported by the project RVO12000.